\documentclass[journal]{IEEEtran}
\ifCLASSINFOpdf
\else
\fi
\usepackage{graphicx}
\usepackage[caption=false]{subfig}
\usepackage{dblfloatfix}    
\usepackage[caption=false]{subfig}
\usepackage{dblfloatfix}
\def\BibTeX{{\rm B\kern-.05em{\sc i\kern-.025em b}\kern-.08em
    T\kern-.1667em\lower.7ex\hbox{E}\kern-.125emX}}
    \usepackage{amsmath}

\begin{document}
%
\title{Design of Beamforming, Transparent Metasurfaces Using Integral Equations}
%
%
%

\author{Malik Almunif, \IEEEmembership{Graduate Student Member, IEEE}, Jordan Budhu, \IEEEmembership{Member, IEEE}, and Anthony Grbic,  \IEEEmembership{Fellow, IEEE}
\thanks{
This work was supported by the UM-KACST Joint Center for Microwave Sensor Technology. }

\thanks{Malik Almunif and Anthony Grbic are with the Department of Electrical and Computer Engineering , University of Michigan, Ann Arbor, MI USA (e-mail: agrbic@umich.edu).}
\thanks{Jordan Budhu is with the Department of Electrical and Computer Engineering, Virginia Tech, Blacksburg, VA, USA (e-mail: jbudhu@vt.edu).}}

\maketitle

\begin{abstract}
An accurate method for designing transmissive metasurfaces is presented that provides perfect transmission while transforming the amplitude and phase of the wavefront. The designed metasurfaces consist of three spatially-varying, electric impedance sheets separated by two dielectric substrates. The design method uses integral equations to account for interactions within and between the impedance sheets, allowing for accurate design. In this paper, a comparison between the integral equation method and the local periodicity approximation is presented. The comparison includes one design example for a transmitted field of uniform phase and amplitude. The design using integral equations provides better collimation. Two other examples involving an amplitude tapered transmitted field are reported to show the versatility of the proposed design technique. In all the examples, the metasurface is $7.35\lambda_0$ wide, the focal length is $4\lambda_0$, and has an overall thickness of $0.1355 \lambda_0 $ at the operating frequency of 5GHz. The designs are verified using a commercial finite element electromagnetic solver.
\end{abstract}

\begin{IEEEkeywords}
Beam shaping, Cascaded sheets, Impedance sheet, Integral equations, Metasurface, Method of Moments
\end{IEEEkeywords}

%
\IEEEpeerreviewmaketitle

\section{Introduction}
\label{sec:introduction}

%
%
%
%
\IEEEPARstart{M}{etasurfaces} are subwavelength-textured surfaces that provide extreme control of wavefronts over very short length scales \cite{b1}--\cite{b3}. They have attracted widespread interest due to the field control they offer, their low profile, and ease of fabrication, which make them suitable for applications across the electromagnetic spectrum \cite{b1,b5}. To date, metasurface designs have been reported that demonstrate a wide range of functions including polarization control, refraction, focusing, and many others \cite{b6}--\cite{b201}. \par

Conventional quasi-optical and optical devices such as lenses are bulky and suffer from reflection at the lens interfaces. To overcome reflections, a metasurface can be used as a matching network \cite{b6,b8,b12,b13}. Metasurface design is usually performed by discretizing the metasurface into small unit cells and designing each unit cell in a periodic environment \cite{b15,b16}, which is referred to as the local periodicity approximation. One major disadvantage of this technique is that it does not account for the presence of dissimilar neighboring unit cells. As a result, authors have used very small spacers \cite{R2_1} between the cascaded-sheets of the metasurface to limit these interactions, leading to additional realization issues and coupling problems.  To mitigate the approximation introduced by the local periodicity assumption in the design and realization of Huygens metasurfaces, baffles were introduced in \cite{b113},\cite{R2_2} to isolate adjacent cells. This method adds complexity to the design and fabrication. In literature, sparse periodic arrangements of scatterers referred to as metagratings, have been used to control the amplitude and phase of wavefronts. However, the design technique is limited to periodic configurations and controlling diffraction orders/grating lobes \cite{n1}--\cite{n3}. More rigorous integral equation modeling and the method of moments (MoM) have been used to design reflective metasurfaces accurately by including the interaction between the adjacent cells \cite{b111,b112}. In \cite{R2_3}, an optimization technique based on MoM and machine learning has been used to design transmissive metasurfaces. However, the optimization is based on the surface parameters, neglecting the intra-cell interaction in the homogenization process.

\begin{figure}[!t]
\centerline{\includegraphics[]{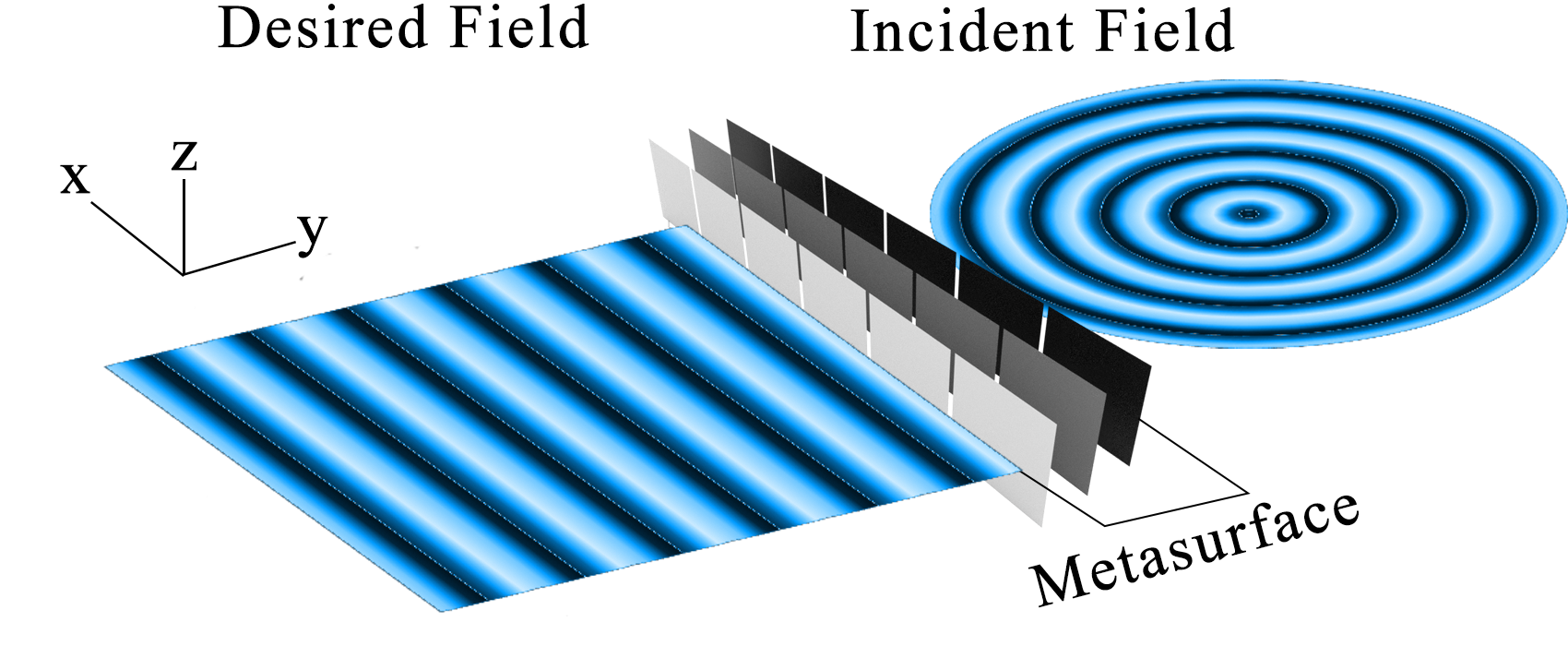}}
\caption{The proposed metasurface geometry. The metasurface consists of three spatially-varying impedance sheets separated by two dielectric substrates. The electromagnetic field of an incident line source is collimated by the metasurface.  The metasurface is invariant in the z-direction.  }
\label{fig1}
\end{figure}

In this paper, we design transparent metasurfaces using integral equations to account for interaction within and between the inhomogeneous sheet impedances. The designed metasurfaces can control both the amplitude and phase of  transmitted fields, enabling the design of a wide range of metasurfaces that beam shape and steer.  Three different examples
are presented in this paper. The metasurfaces consist of three spatially-varying, cascaded sheets separated by two dielectric substrates, as shown in Fig. \ref{fig1}.  Full-wave simulations are performed using Ansys HFSS with a line source excitation to verify the results obtained through integral equations.  \par
The paper is organized as follows, Section \ref{sec2} describes the conventional technique for designing metasurfaces where each unit cell is simulated with periodic boundaries.  Section \ref{sec3} outlines the integral equation design technique that utilizes optimization. It also compares the two design methods. Section \ref{sec4} presents two design examples showing the capability of the integral equation method. The performance of the metasurfaces is verified using a commercial full-wave solver.

\section{Metasurface Lens Design using Local Periodicity}
\label{sec2}
\begin{figure}[!t]
\centerline{\includegraphics[]{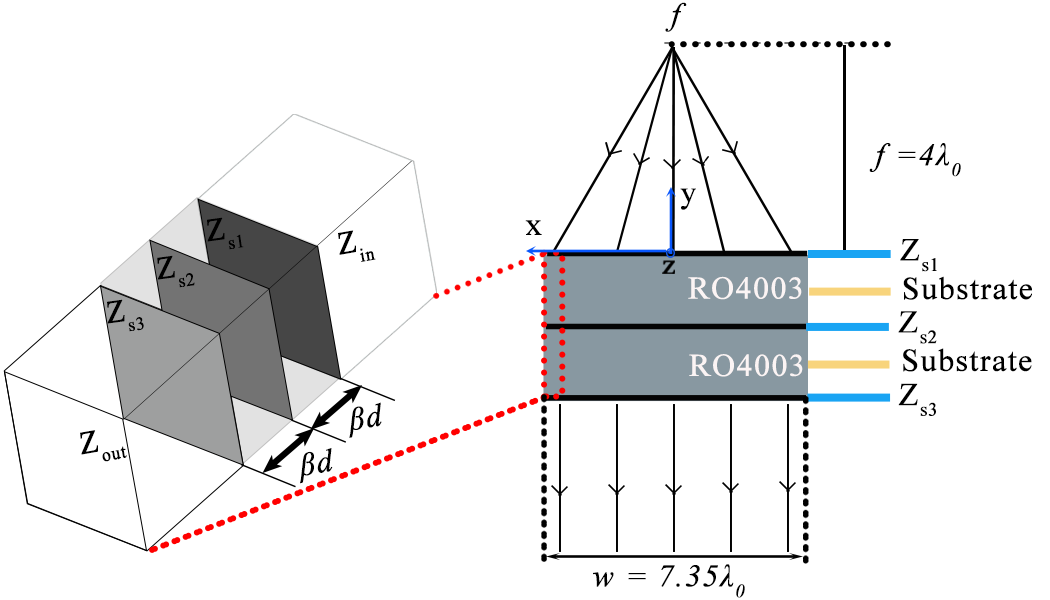}}
\caption{Transmissive metasurface design using local periodicity. Three cascaded sheets are separated by an electrical separation $\beta d$ ($\beta$ is the propagation constant). The inset shows the three constitutive homogeneous sheets of a unit cell.  }
\label{fig2}
\end{figure}

In this section, a metasurface that collimates the field of a line source is designed using the local periodicity approximation. The metasurface is discretized into electrically small unit cells. Each unit cell is composed of three cascaded homogeneous sheet impedances, as seen in Fig. \ref{fig2}. Given the local TE wave impedances of the incident and transmitted waves, and the transmission phase through each unit cell, the desired local $Z$ (impedance) parameters ($Z_{11}$, $Z_{12}=Z_{21}$, $Z_{22}$) of each reflectionless unit cell are found. From these $Z$ parameters, the three homogeneous sheet impedance values needed for each unit cell are computed using the following expressions (see \cite{b16}, for the detailed procedure),
\newcounter{storeeqcounter}
\newcounter{tempeqcounter}

\begin{align}
 Z_{s1} ={}& \frac{jZ_0\sin{(2\beta d)} - \frac{Z^2_0}{Z_{s2}}\sin^2{(\beta d)}}{\frac{Z_{22}}{Z_{21}} - 1 + 2\sin^2{(\beta d)} - j\frac{Z_0}{2Z_{s2}} \sin{(2\beta d)}}\\
  Z_{s2} ={}& \frac{Z^2_0\sin^2{(\beta d)}}{jZ_0\sin{(2\beta d)} - \frac{\left|Z\right|}{Z_{21}}}\\
   Z_{s3}={}& \frac{jZ_0 \sin{(2\beta d)} - \frac{Z_0^2}{Z_{s2}}\sin^2{(\beta d)}}{\frac{Z_{11}}{Z_{21}} - 1 + 2\sin^2{(\beta d)}-j\frac{Z_0}{2Z_{s2}}\sin{(2\beta d)}}
\end{align}

\noindent
where $Z_0$ is the wave impedance of the dielectric substrate and $\beta d$ is the electrical separation in radians between the impedance sheets. These equations assume that the three sheet impedances are periodic, that is they are invariant along the transverse axis (x-axis).   \par

 Each unit cell comprises three homogeneous sheet impedances that provide three degrees of freedom. Two degrees of freedom are needed to locally match the impedance of the incident wave to that of the transmitted wave, while the third controls the transmission phase through the unit cell. For a metasurface that collimates the field radiated by a line source, the output impedance of each unit cell is the wave impedance of free space $Z_{out}= \eta_0 = 377\Omega$. The input impedance for each unit cell (shown in Fig. \ref{fig2}) depends on the local angle of incidence. The angle of incidence is computed from the geometry as $\theta = \sin^{-1}(1-\frac{f}{\sqrt{f^2 +x^2_n}})$, where $f$ is the focal length and $x_n$ is the distance from origin. Thus, the local input impedance is the TE wave impedance $Z_{in}= \frac{\eta_0}{\cos{\theta}}$. A quadratic phase distribution $\phi_n = \phi_0 + k_0(\sqrt{f^2 + x_n^2}-f)$ is impressed by the metasurface, where $\phi_n$ is the phase shift at $x_n$, $\phi_0$ is a common phase shift, and $k_0$ is the wave vector.\par

Each unit cell is designed assuming two RO4003C substrates (4.065 mm thick with $\epsilon_r = 3.55$) are placed between the impedance sheets. The focal length for the design is chosen to be at $4\lambda_0$, the diameter of the metasurface is $7.35 \lambda_0$, and the unit cell width is $\frac{\lambda_0}{20}$. The separation between the impedance sheets is set to $\frac{\lambda_0}{14.76}$ for an operating frequency of 5 GHz.  The sheet impedances calculated for each unit cell using this method assume the cells are in a periodic environment. This assumption results in inaccurate mutual coupling modeling of the metasurface given that it is aperiodic.

\section{Metasurface Design using Integral Equations}
\label{sec3}

In this section, the metasurface is accurately modeled using integral equations which account for the interactions within and between the inhomogeneous impedance sheets. In design, the sheet impedances comprises the metasurface are optimized using integral equation modeling to provide the desired transmitted field. It is shown that the earlier local periodicity approach results in an inaccurate design as compared to the integral equation modeling approach. This was pointed out in \cite{b111} for reflection, here, we consider the transmission case. 
\subsection{The Integral Equation Formulation}
The total electric field along each layer is given by,
\begin{equation}
E_{tot} = E_{inc} + E_s
\label{eq4}
\end{equation}
where $E_{inc}$ is the incident field, and $E_s$ is the field scattered by the current density on the sheets and the substrate. The total electric field $E_{tot}$ can be related to the current density by the boundary condition $E_{tot} = Z_{si}J_{zi}$ at the $ith$ sheet where $i=1,2,3$. The incident field is a z-directed TE polarized line source with a constant current $I_0$, 

\begin{equation}
    E_{inc} = \frac{-k_0 \eta_0 I_0}{4} H_0^{(2)} (k_0\sqrt{x^2+(y-f)^2})
\end{equation}
where $H_0^{(2)}$ is the Hankel function of the second kind of order zero. The field scattered onto the first sheet impedance is given by, 
\begin{multline}
         E_{s1}  =   -j\eta_0 k_0 \Bigg[ \displaystyle \sum_{i=1}^{3}  \int_{\frac{-w}{2}}^{\frac{w}{2}} J_{zi} G(x,0,x^\prime,(1-i)d)dx^\prime  
\\+   \displaystyle \sum_{k=4}^{5}  \int_{\frac{-d_k}{2}}^{\frac{d_k}{2}} \int_{\frac{-w}{2}}^{\frac{w}{2}} J_{zk} G(x,0,x^\prime,y^\prime) dx^\prime dy^\prime
 \Bigg]
 \label{eqn_sheet}
\end{multline}
where $J_{z1}, J_{z2}$, and $J_{z3}$ represent the current density on the three cascaded sheet impedances, and the polarization current density induced within the substrates are labeled as $J_{z4}$ and $J_{z5}$.  The field scattered onto the first substrate layer is given by, 
\begin{multline}
     E_{s4}  =    -j\eta_0 k_0 \Bigg[ \displaystyle \sum_{i=1}^{3}  \int_{\frac{-w}{2}}^{\frac{w}{2}} J_{zi} G(x,y,x^\prime,(1-i)d)dx^\prime  
\\+   \displaystyle \sum_{k=4}^{5}  \int_{\frac{-d_k}{2}}^{\frac{d_k}{2}} \int_{\frac{-w}{2}}^{\frac{w}{2}} J_{zk} G(x,y,x^\prime,y^\prime) dx^\prime dy^\prime
 \Bigg]  
 \label{eqn_sub} 
\end{multline}

In (\ref{eqn_sheet}) and (\ref{eqn_sub}), prime coordinates are used for the location of sources and unprimed coordinates for the observation locations, $w$ is the metasurface width, $d$ is the thickness of both dielectrics. A similar procedure can be applied to find the scattered field onto each remaining layer.  The 2D Green's function of free space appearing in (\ref{eqn_sheet}) and (\ref{eqn_sub}) is, 

\begin{equation}
    G(x,y,x^\prime,y^\prime) = \frac{1}{4j}H_0^{(2)}(k_0\sqrt{(x-x^\prime)^2+(y-y^\prime)^2}
\end{equation}

To solve the integral equation, the MoM can be used \cite{b111,b112,b114}. The surface current density on each sheet impedance is discretized into N pulse basis functions. The number of pulses is chosen to adequately capture the variation in current along the sheet. The variation in current depends on the range of impedances used in the metasurface design. Reactances that are closer to zero require more pulses since they support currents with larger transverse wavenumbers. In the designs presented, the current on each homogeneous sheet impedance is discretized into 14 pulses. Therefore,  the total number of pulses N on each sheet impedance is 2058. A similar procedure is used to model the volume current density in the substrate, but with 2D pulses. In the examples presented, 5 pulses are used in the y-direction and 500 pulses in the x-direction. Applying Galerkin testing to the equations, results in a linear system that can be written in a matrix form,

\begin{equation}
    \begin{bmatrix}
[V]
\end{bmatrix}
= 
\begin{bmatrix}
[Z]+[Z_{s}] \end{bmatrix}
\begin{bmatrix}
[I]
\end{bmatrix}
\label{MoM}
\end{equation}
where $[V]$ is a vector containing the field incident onto each layer, $[Z]$ is the mutual coupling matrix both within and between each layer, $[Z_s]$ is a diagonal matrix of sheet and substrate impedances, and $[I]$ is the current density vector. For the definitions of matrices in (\ref{MoM}), see the appendix of \cite{b112}.

\subsection{Optimization}
Once the current density is calculated through matrix inversion, the resulting electric field can be calculated using (\ref{eq4}). The analytical values calculated in Section \ref{sec2} are used as a seed for the optimization. The optimization cost function is defined at a reference plane one wavelength away from the metasurface. In this first metasurface design, the goal is to reshape the incident line source into a uniform phase and amplitude without reflections at the interface. The source is placed at a distance $f = 4\lambda_0$ in front of the metasurface. Thus, to ensure full transmission, the incident power is equated to the transmitted power $P_i = P_t$, where the transmitted power per unit length is $P_t = \frac{|E_{des}|^2}{2\eta_0}w$. Hence, the desired electric field magnitude per unit length can be written as, 
\begin{equation}
    |E_{des}| = \sqrt{\frac{2\eta_0 P_i}{w}}
\end{equation}
In addition, a uniform phase can be achieved by setting the phase at the optimization reference plane equal to a constant, arbitrary phase. Therefore, the cost function for the optimization is $ f = |E_{tot} - E_{des}|$

\begin{figure}[!t]
\subfloat[]{\includegraphics[]{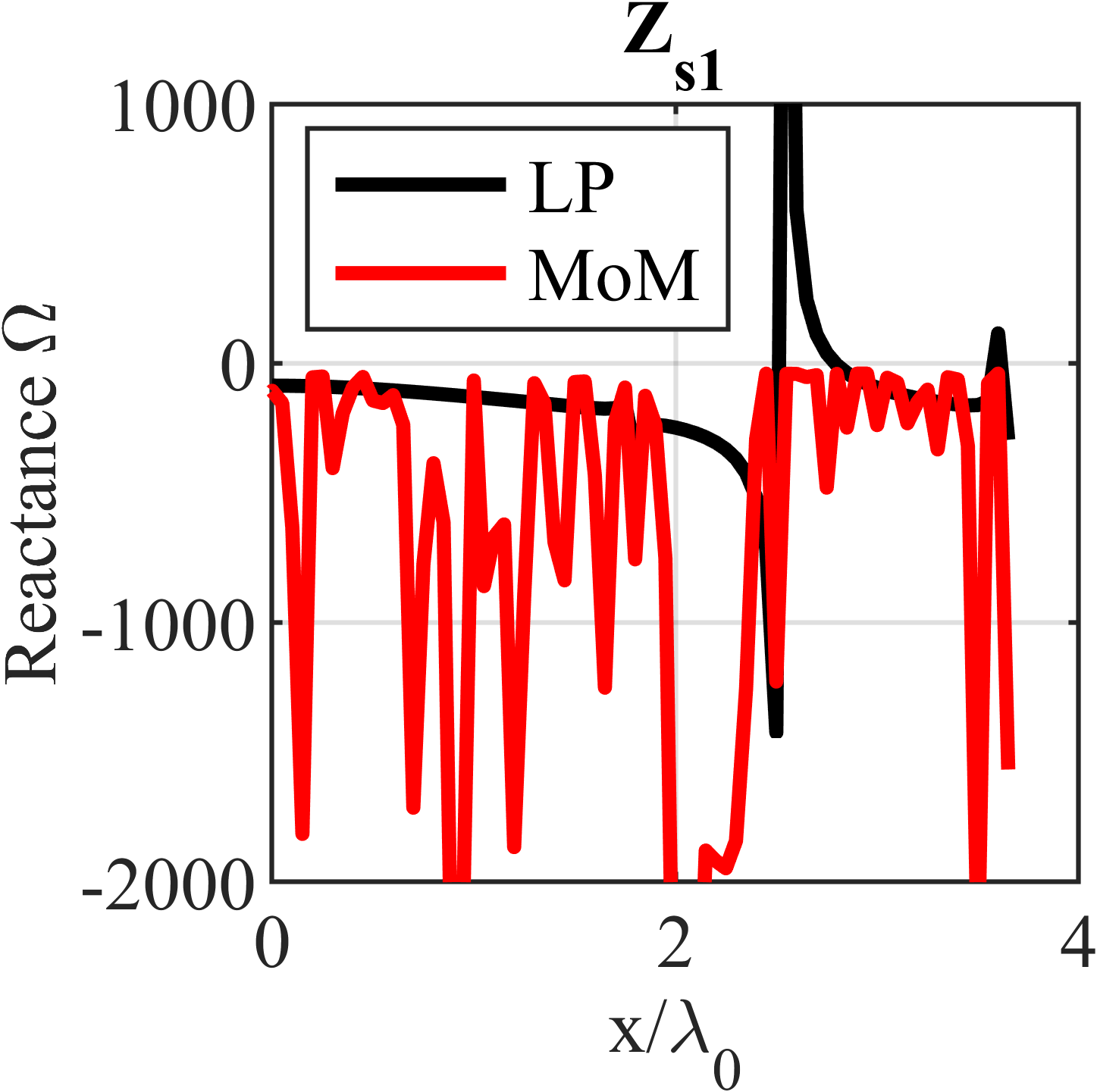}}\hfill
	\subfloat[]{\includegraphics[]{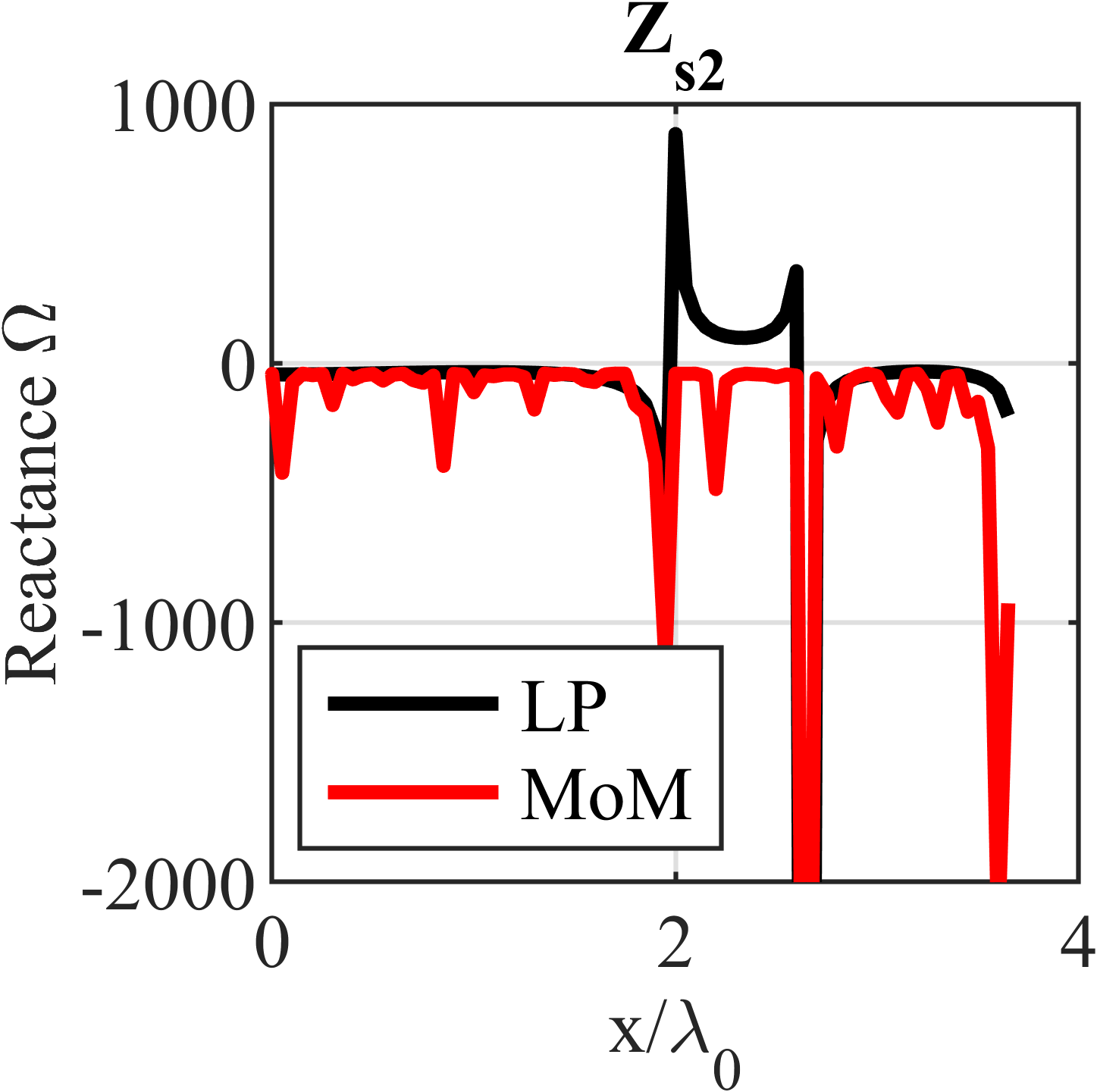}}\\
	\centering
		\subfloat[]{\includegraphics[]{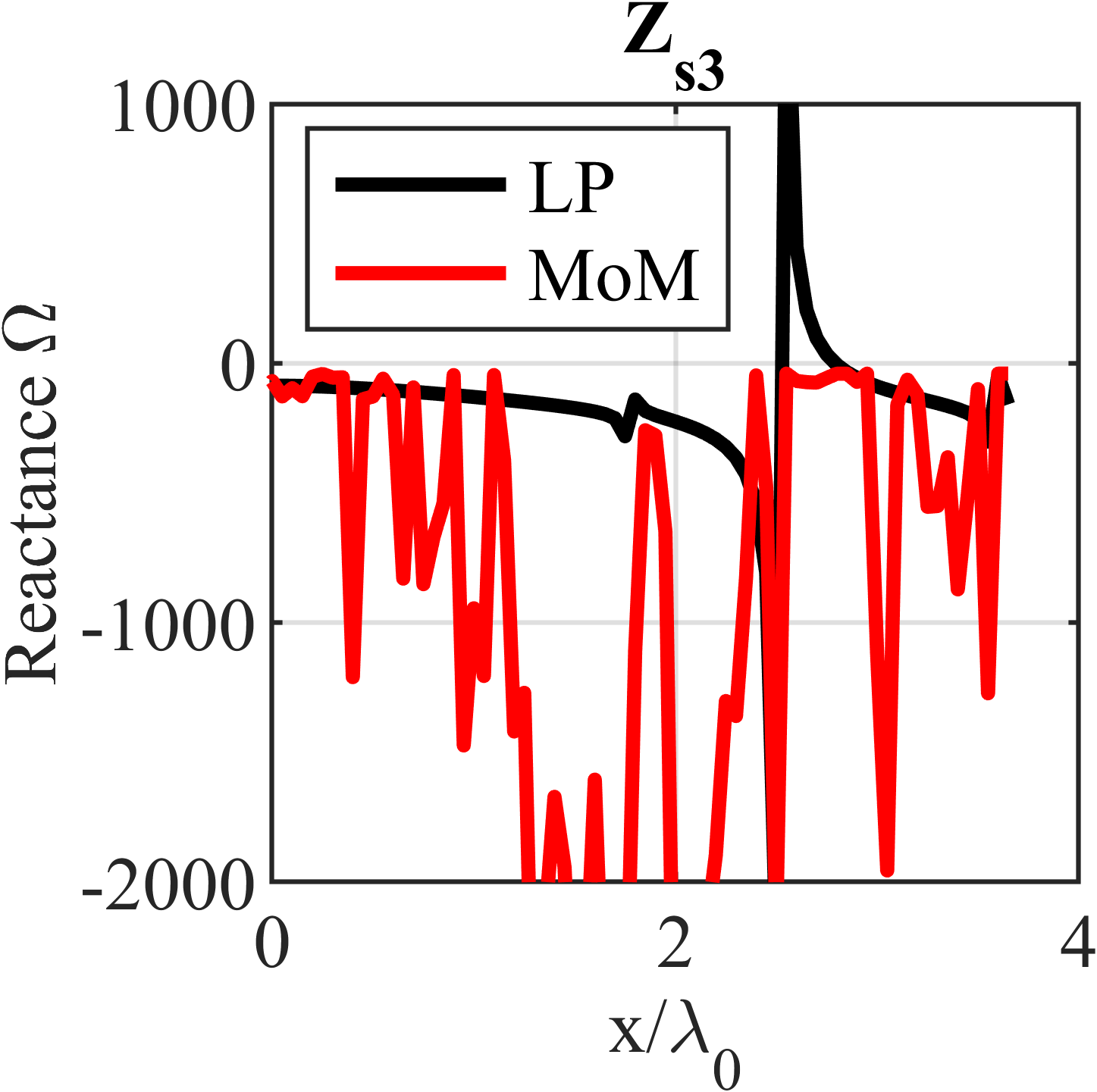}}\\	
	\caption{The impedance profiles needed to implement the uniform phase and amplitude metasurface using local periodicity approach and integral equation based optimization approach at an operating frequency 5 GHz: (a) First layer $Z_{s1}$. (b) Second layer $Z_{s2}$. (c) Third layer $Z_{s3}$. Half the impedance profile is shown due to symmetry.}
	\label{fig3}
\end{figure}

The MATLAB function fmincon is used to minimize the cost function \cite{b115},\cite{R2_5}. To generate a purely reactive impedance profile, the optimization adds rapid changes in the impedance profile which introduces surface waves \cite{b202}, \cite{b300}.  In the optimization, we have used capacitive impedances because they support TE surface waves \cite{nn1}. With increasing capacitance, the transverse wavenumbers of supported surface waves become larger. Thus, to limit the minimum transverse wavelength of the supported surface waves to twice the unit cell dimension,  the sheet reactances used in the optimization must be less than $-j20 \Omega$ \cite{b202}.   For a cell size equal to $\lambda/20$, this limited the extent of the evanescent spectrum to surface waves with a transverse wavenumber of $10k_0$. To be conservative, the sheet reactances are set to be less than $-j40\Omega$. 
 The optimized sheet impedances are compared to the impedances obtained using the local periodicity approximation in Fig. \ref{fig3}. The optimized sheet impedances generally follow the same envelope as the sheet impedances calculated using local periodicity.

\subsection{Comparison between Local Periodicity and MoM Design Procedures}

\begin{figure}[!t]
\centering \subfloat[]{\includegraphics[scale=0.95]{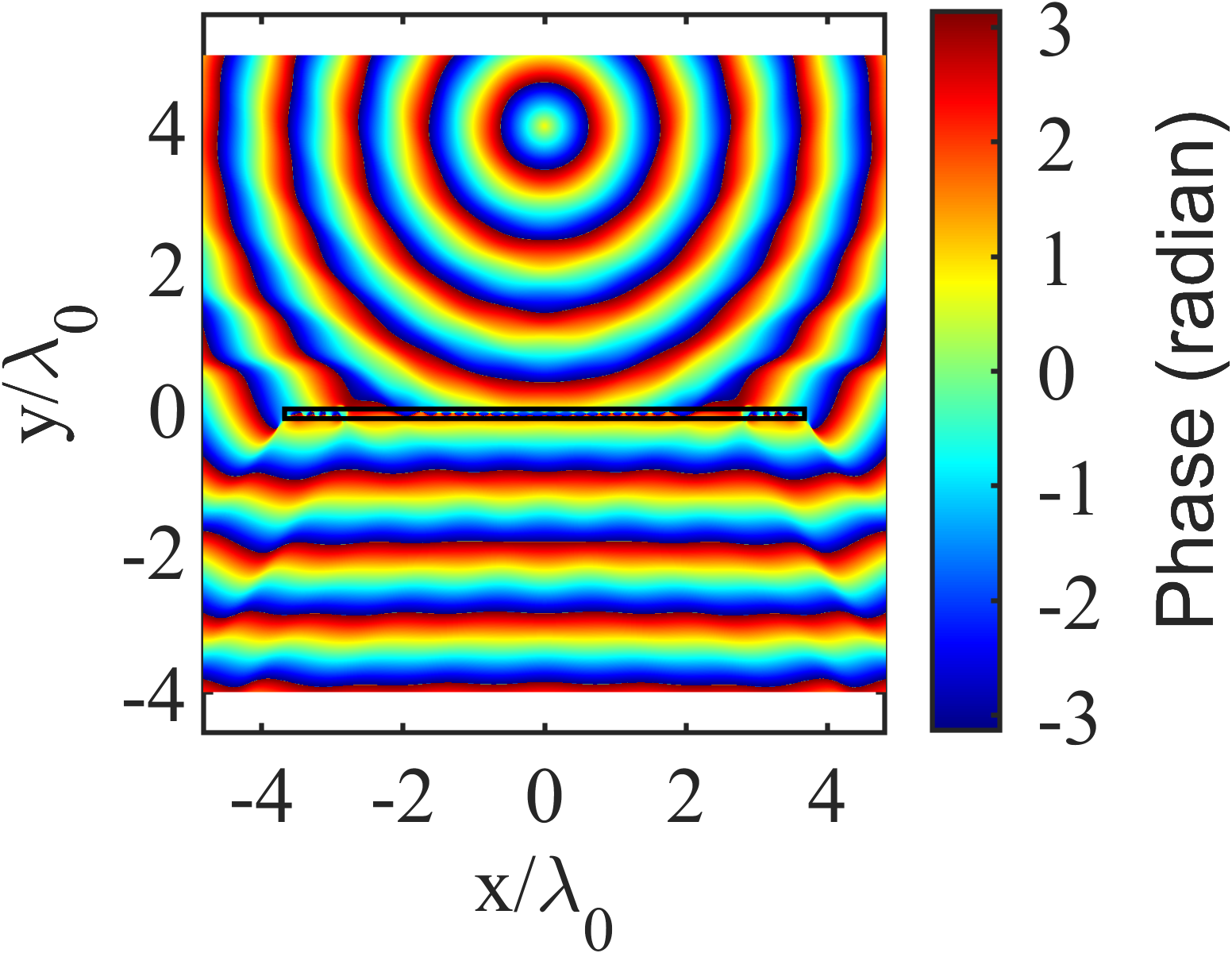}}\hfill
	\subfloat[]{\includegraphics[scale=0.95]{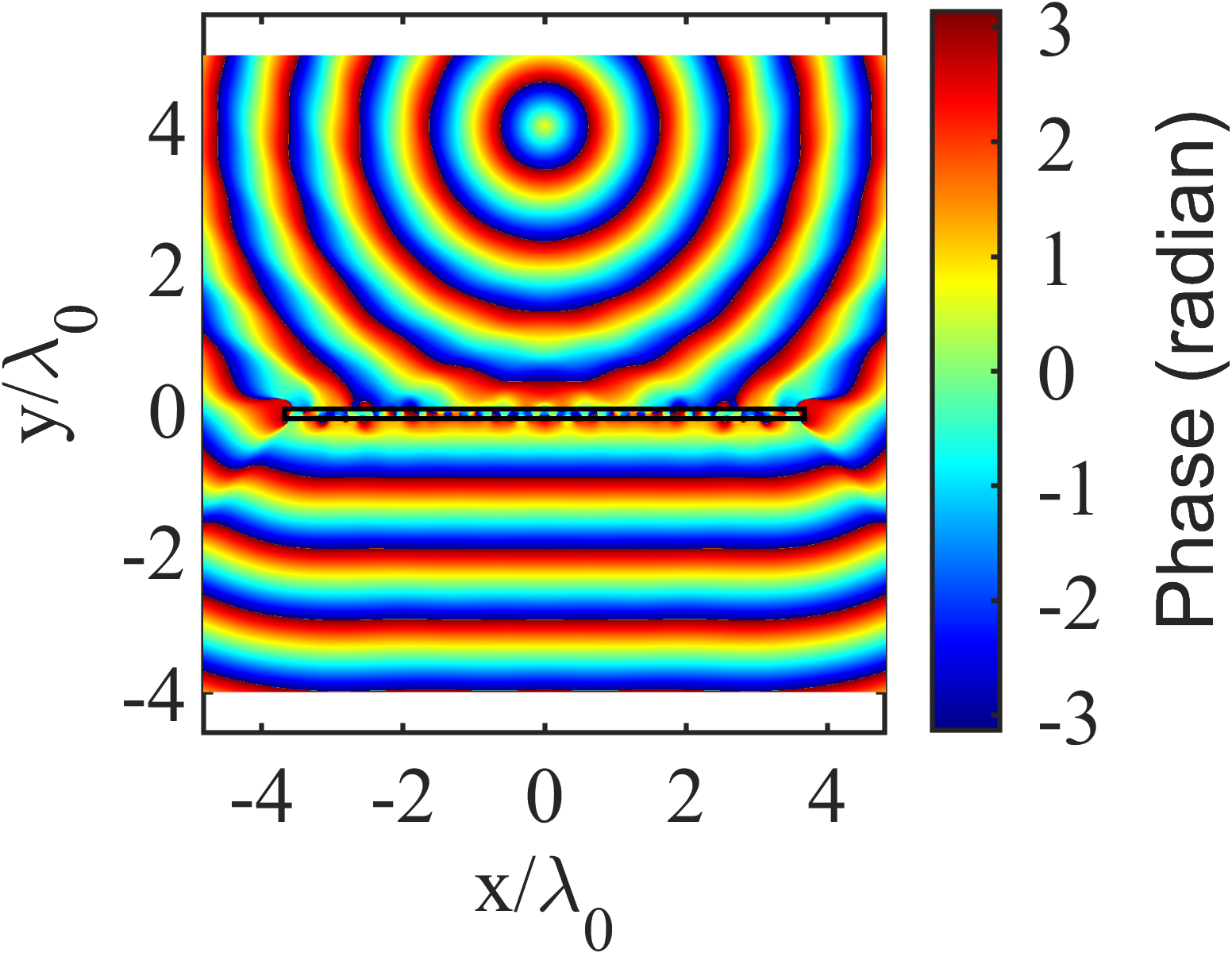}}\\
	\caption{The electric field phase plot obtained from Ansys HFSS for the two metasurface design approaches. (a) Local periodicity design method. (b) MoM with optimization design method.  }
	\label{fig4}
\end{figure}

\begin{figure}[!t]
\centerline{\includegraphics[]{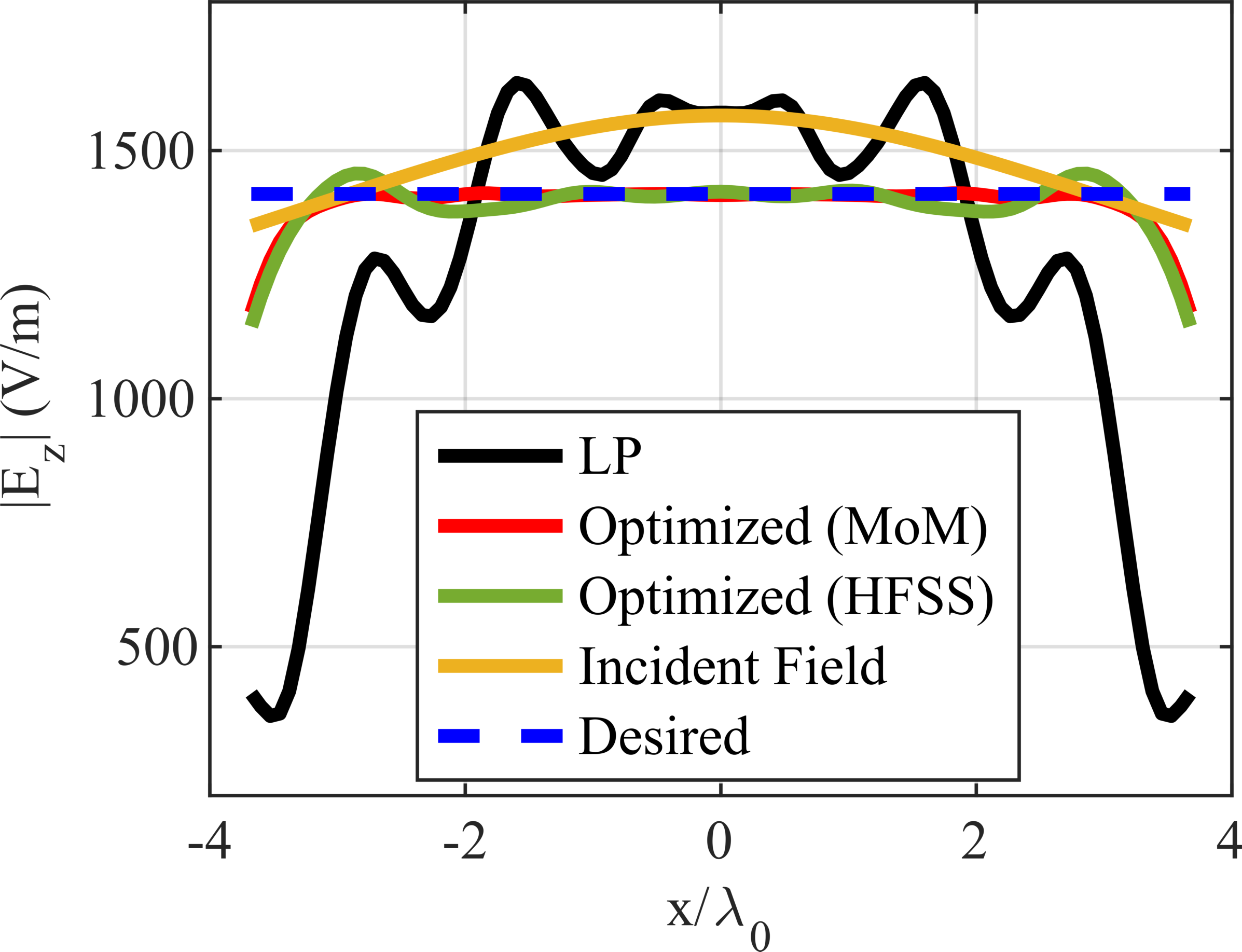}}
\caption{A comparison of the metasurfaces designed using the local periodicity method and MoM with optimization method. The results are verified using Ansys HFSS. The electric field magnitude plotted one wavelength away from the metasurface.  }
\label{fig5}
\end{figure}

A comparison between designs obtained through the local periodicity approximation and integral equation approaches is detailed in this subsection. Fig. \ref{fig4}(a) shows a top view of the electric field phase resulting from a line source excitation incident onto the metasurface composed of ideal sheet impedances designed using the local periodicity method. Replacing these sheets with the optimized sheets from the integral equation method produces more accurate results in terms of uniform phase and amplitude, as shown in Fig. \ref{fig4}(b). To compare the two methods, the electric field magnitude is plotted along a surface parallel to the xz plane, as shown in Fig. \ref{fig5}. Full-wave verification of the results is also shown using the commercial electromagnetics solver Ansys HFSS.  \par The results obtained using the MoM show that the metasurface is capable of controlling the phase and amplitude of the transmitted field. Using integral equations, the transmitted field can be reshaped to any desired amplitude and phase. 

\subsection{Bandwidth}
\begin{figure}[!t]
\subfloat[]{\includegraphics[]{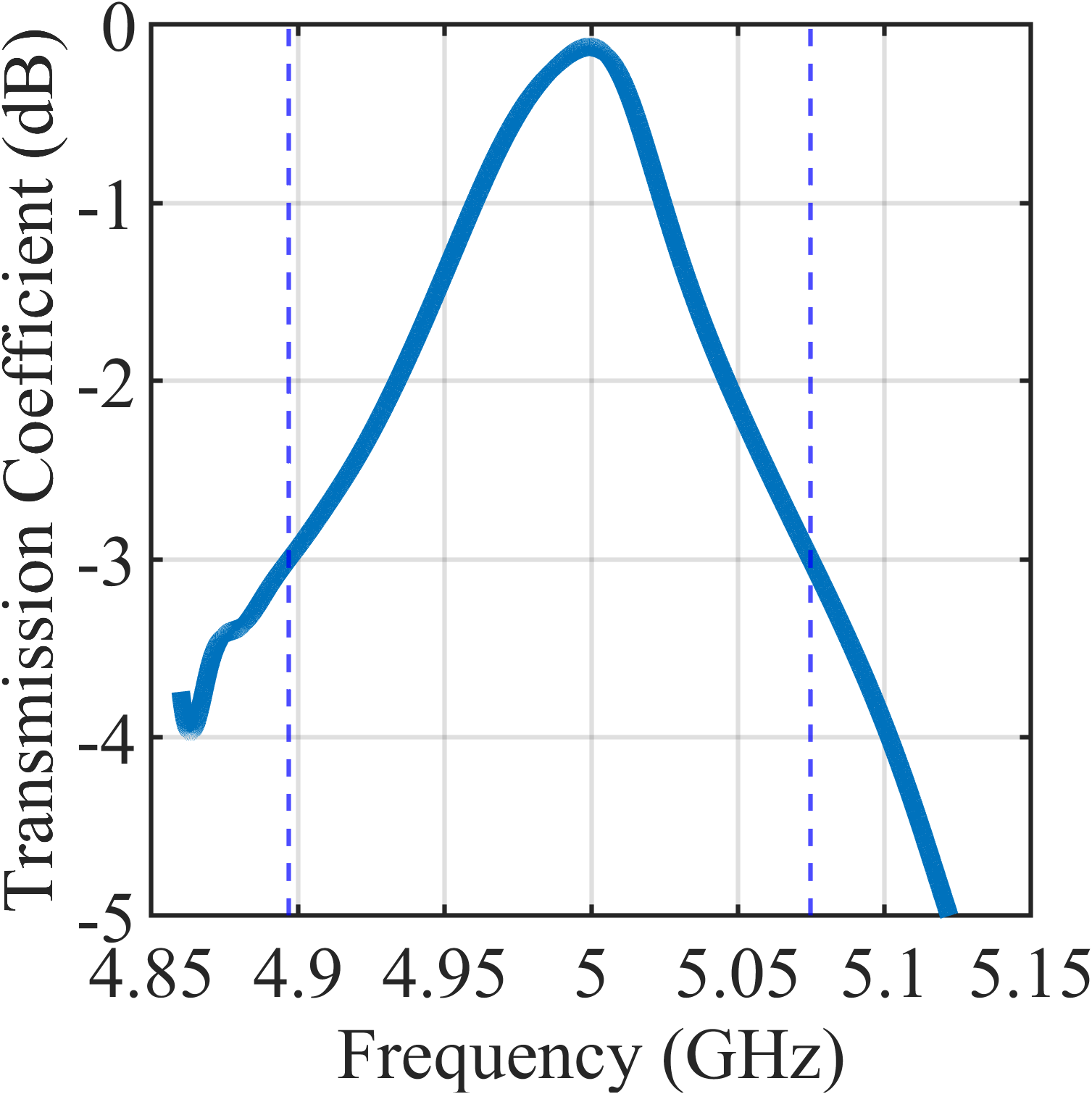}}\hfill
	\subfloat[]{\includegraphics[]{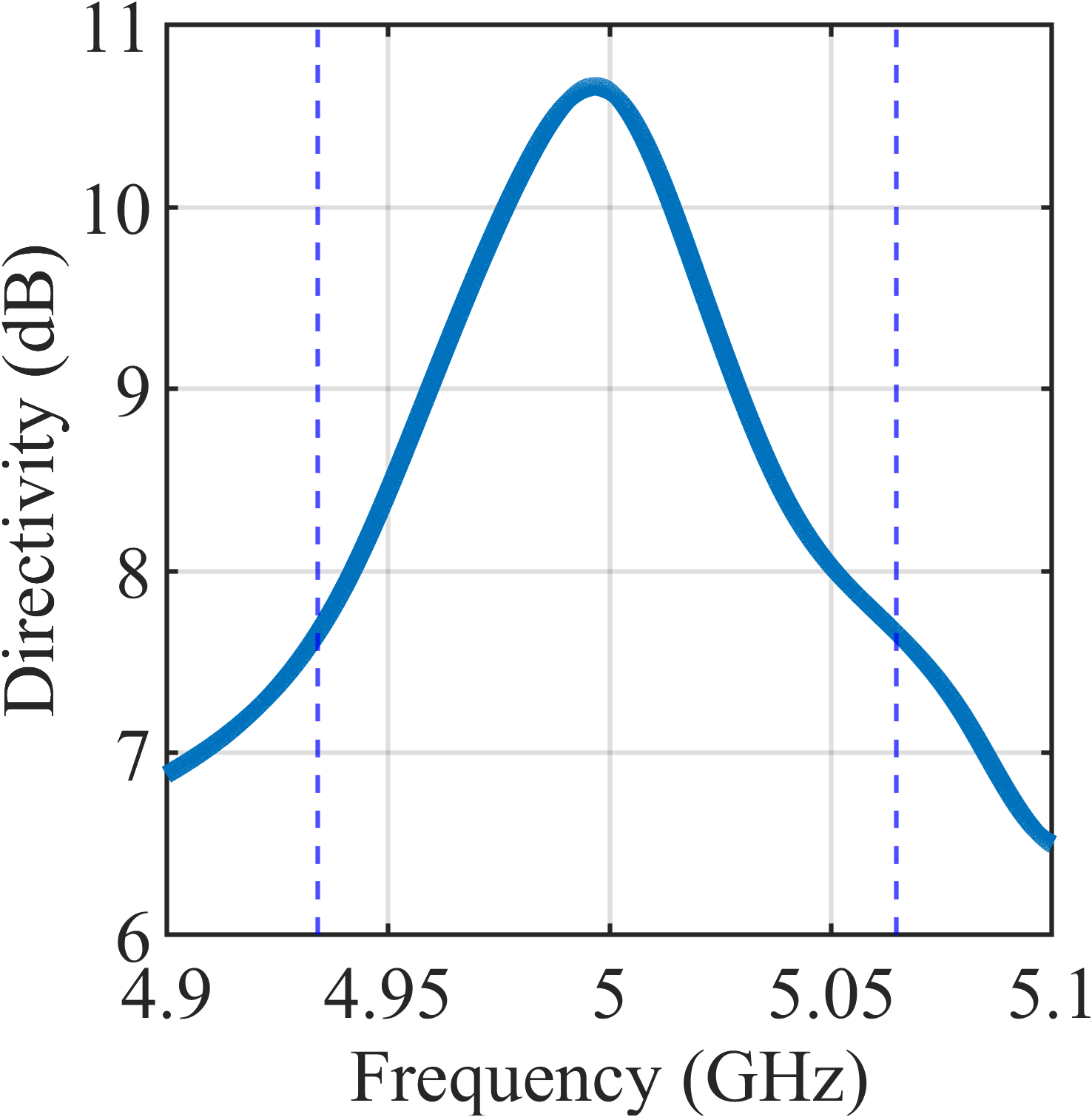}}\\
	\caption{Bandwidth extracted from Ansys HFSS for the integral equation-based design: a) 3dB transmission bandwidth. b) 3dB directivity bandwidth.   }
	\label{figBW}
\end{figure}

The bandwidth for the first example designed using integral equations is calculated using the commercial full-wave solver Ansys HFSS. The capacitive impedance profile is scaled for each frequency by $f_0/freq$, where $f_0$ is the frequency of operation 5GHz. The 3dB transmission bandwidth is calculated to be 3.56\%, as shown in Fig.  \ref{figBW}(a).  Where the transmission coefficient is defined as $T = 10\log(P_t/P_i)$. In addition, we have calculated the 3dB directivity bandwidth to be 2.61\%, as shown in Fig. \ref{figBW}(b). 
\section{Amplitude and Phase Control for Beam Shaping}
\label{sec4}

In this section, we describe two design examples where both the phase and amplitude of the transmitted field are controlled using the proposed technique. The first example involves collimating and Gaussian tapering the field radiated by a line source. The second example involves collimating and scanning the field radiated by a line source to a 35 degree angle.  The same substrate and discretization as in the previous example are used. 

\subsection{Metasurface Design Example II: Gaussian Tapering}

\begin{figure}[!t]
\centerline{\includegraphics[]{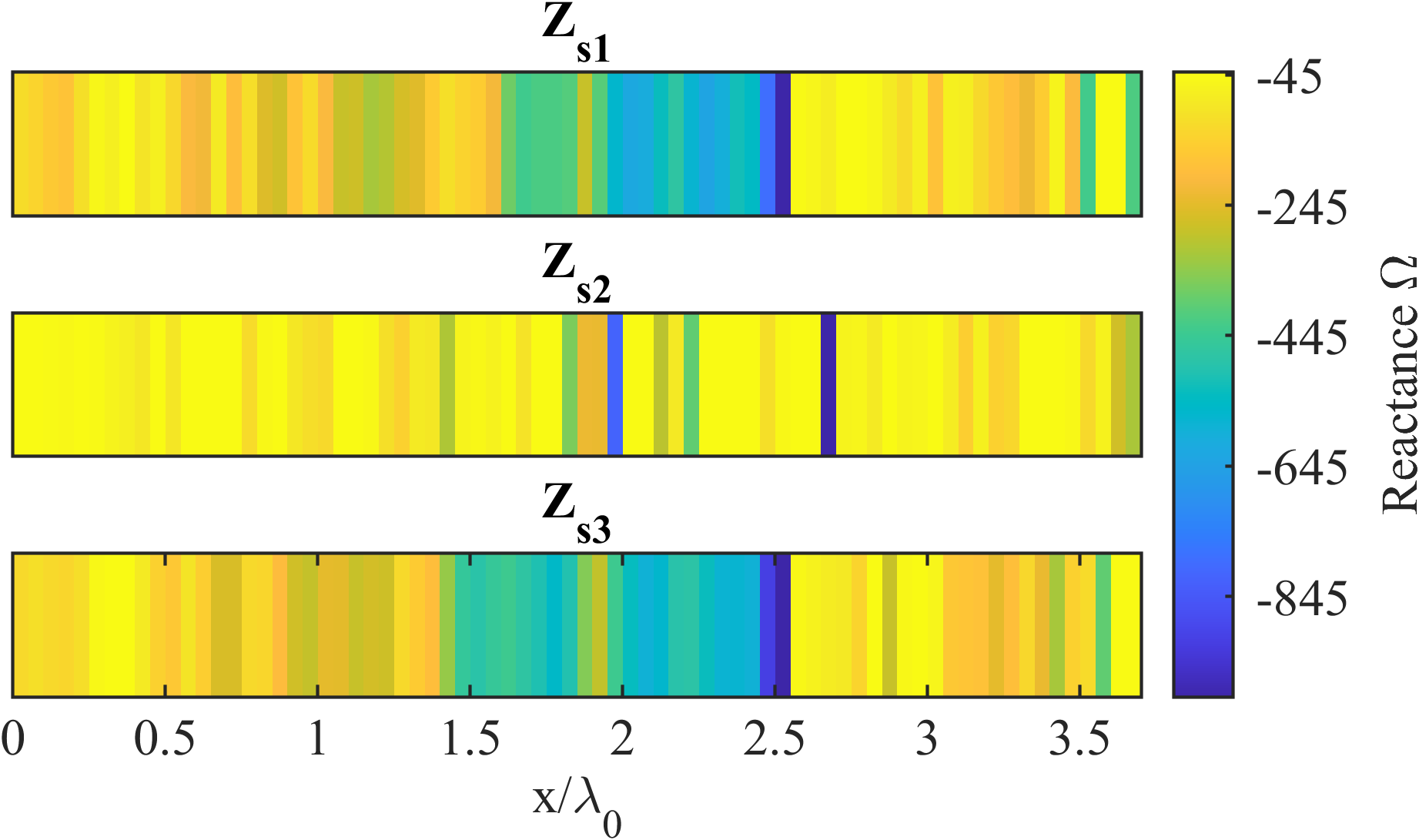}}
\caption{The impedance profiles for the metasurface designed to implement a Gaussian taper of the transmitted collimated field. Since the metasurface is symmetric, only half of the impedance profile is shown in the figure.    }
\label{fig6}
\end{figure}

The transmitted field is assumed to have a Gaussian taper of the following form along the optimization plane,

\begin{equation}
    E_{z} = E_0 e^{-(\frac{2}{w}x)^2}
\end{equation}
where $w$ is the metasurface width. The amplitude $E_0$ is selected to ensure full transmission and a lossless aperture. The electric field is Fourier transformed in order to calculate the transmitted magnetic field spectrum: $    \Tilde{H}_x = \frac{\Tilde{E}_z k_y}{\eta_0 k_0}$, where $\Tilde{}$ defines a spectral quantity. By inverse Fourier transforming the magnetic field spectrum, we can find the total transmitted power per unit length as $P_t = \frac{1}{2} \int_{-w/2}^{w/2}  Re[E\times H^*] dx$. Thus, the electric field magnitude $E_0$ can be found by scaling the input power to the transmitted power, 
\begin{equation}
\label{E_0}
    E_0 = \sqrt{\frac{P_i}{P_t}}
\end{equation}
After following the procedure outlined in the previous section, we found the required sheet impedance to collimate and  Gaussian amplitude taper the transmitted field. The impedance profiles used to realize the Gaussian taper and collimation is shown in Fig. \ref{fig6}.

\begin{figure}[!t]
\subfloat[]{\includegraphics[]{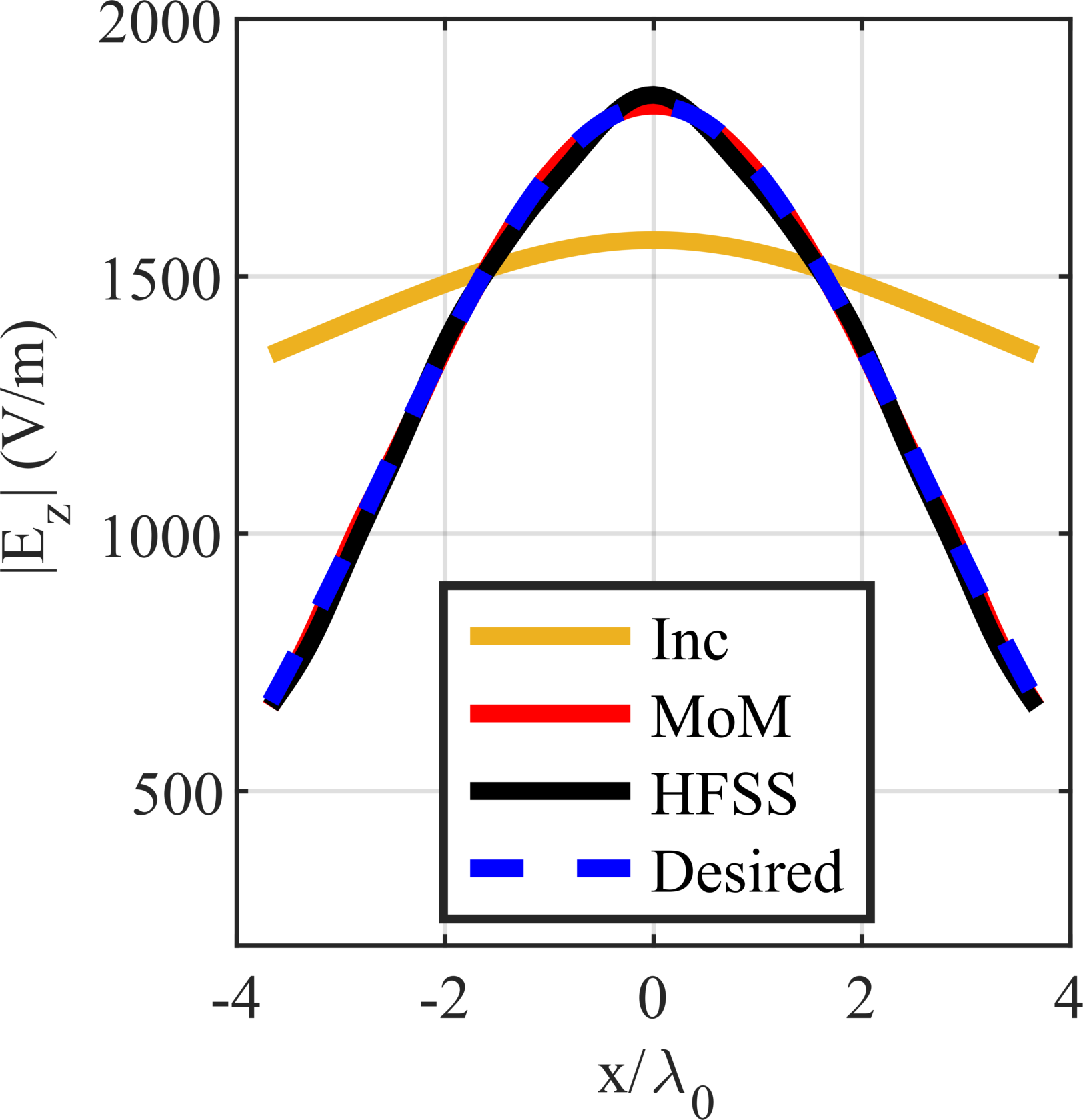}}\hfill
	\subfloat[]{\includegraphics[]{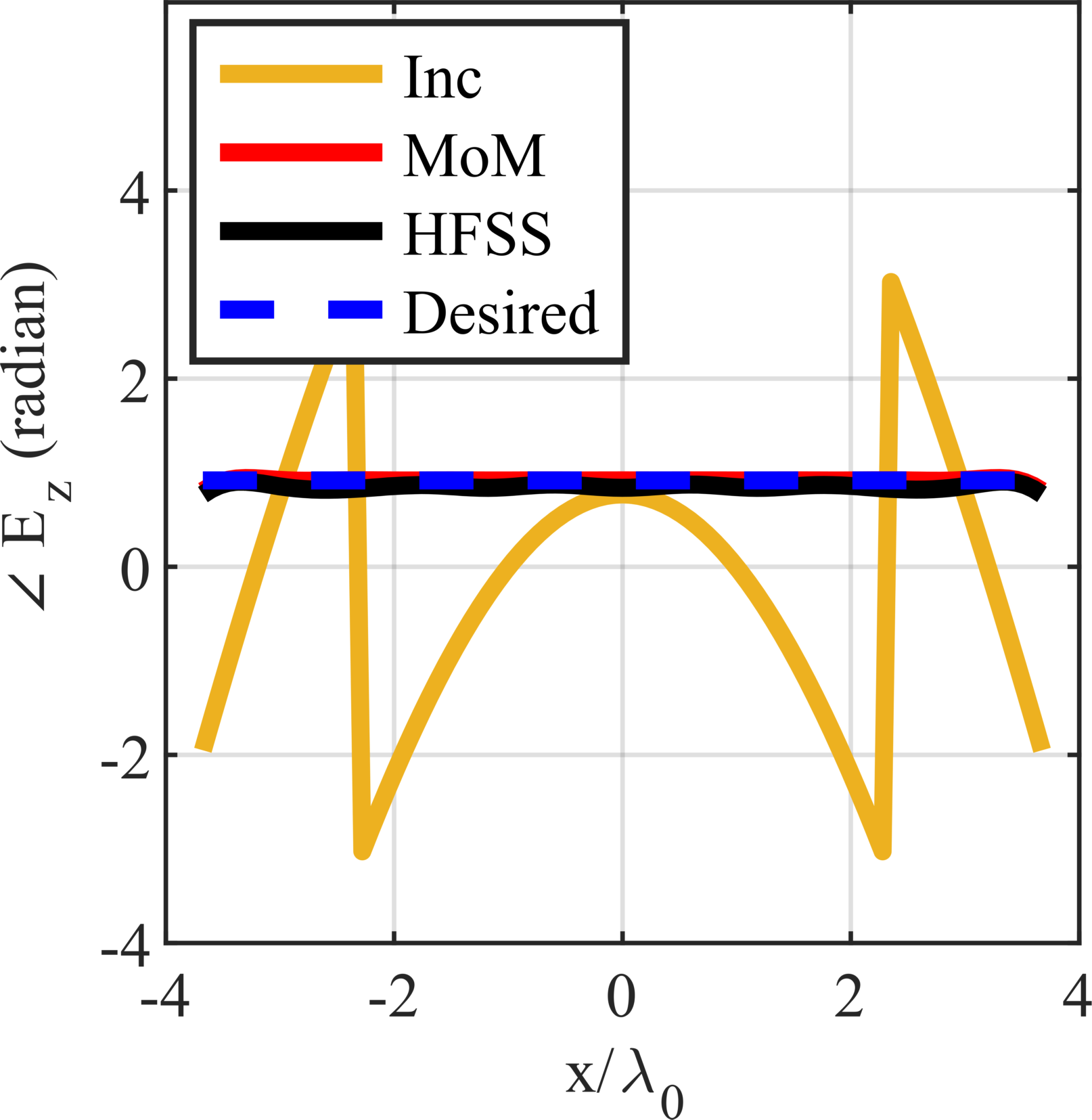}}\\
	\caption{The total electric field transmitted by the metasurface for example II. The incident field is shown in yellow, the desired field in dashed blue, the transmitted field computed using MoM in red, and that computed using HFSS is in black. (a) Electric field amplitude. (b) Electric field phase.  }
	\label{fig7}
\end{figure}

\begin{figure}[!t]
\centerline{\includegraphics[scale=1]{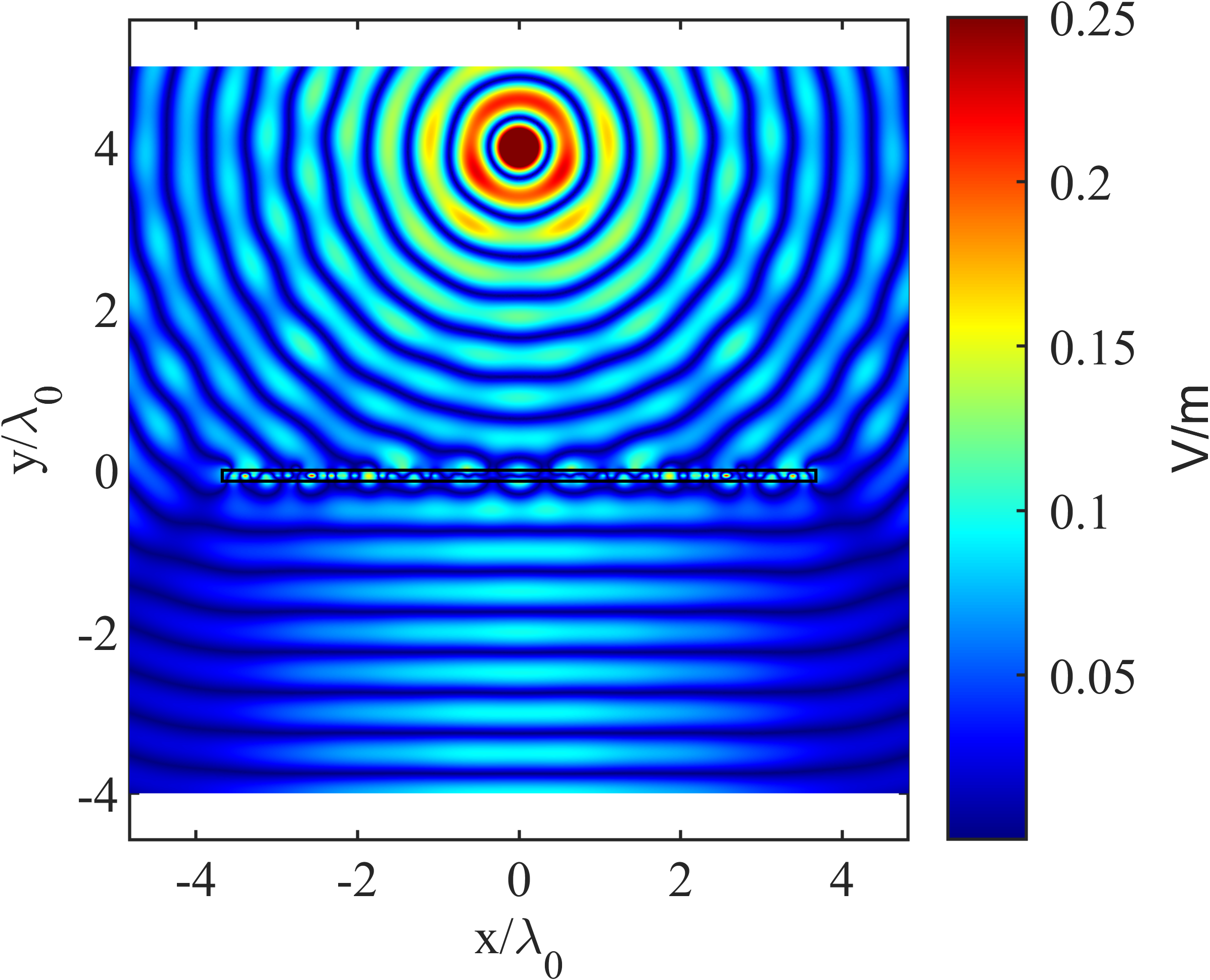}}
\caption{A time snapshot of the electric field from an HFSS full wave simulation for example II. The electric field is normalized to 25\% of the feed maximum value.   }
\label{fig8}
\end{figure}

Fig. \ref{fig7} shows the complex electric field magnitude and phase on a plane that is $\lambda_0$ away from the metasurface. The plot shows that the metasurface produces an amplitude distribution that closely agrees with the desired field.  Also, the results are verified through full-wave simulation using Ansys HFSS. A time snapshot of the electric field from Ansys HFSS full-wave simulation is shown in Fig. \ref{fig8}.

\subsection{Metasurface Design Example III: Scanned Beam}
In the third metasurface design, the electric field amplitude is cosine tapered and scanned to an angle of 35 degree, while the incident field is kept the same. The phase of the desired electric field is achieved by applying a phase gradient that supports the desired scan angle. The electric field amplitude for the cosine taper is given by, 

\begin{equation}
    E_z = E_0 \cos{(\frac{\pi}{w}x)}
\end{equation}

\begin{figure}[!t]
\centerline{\includegraphics[]{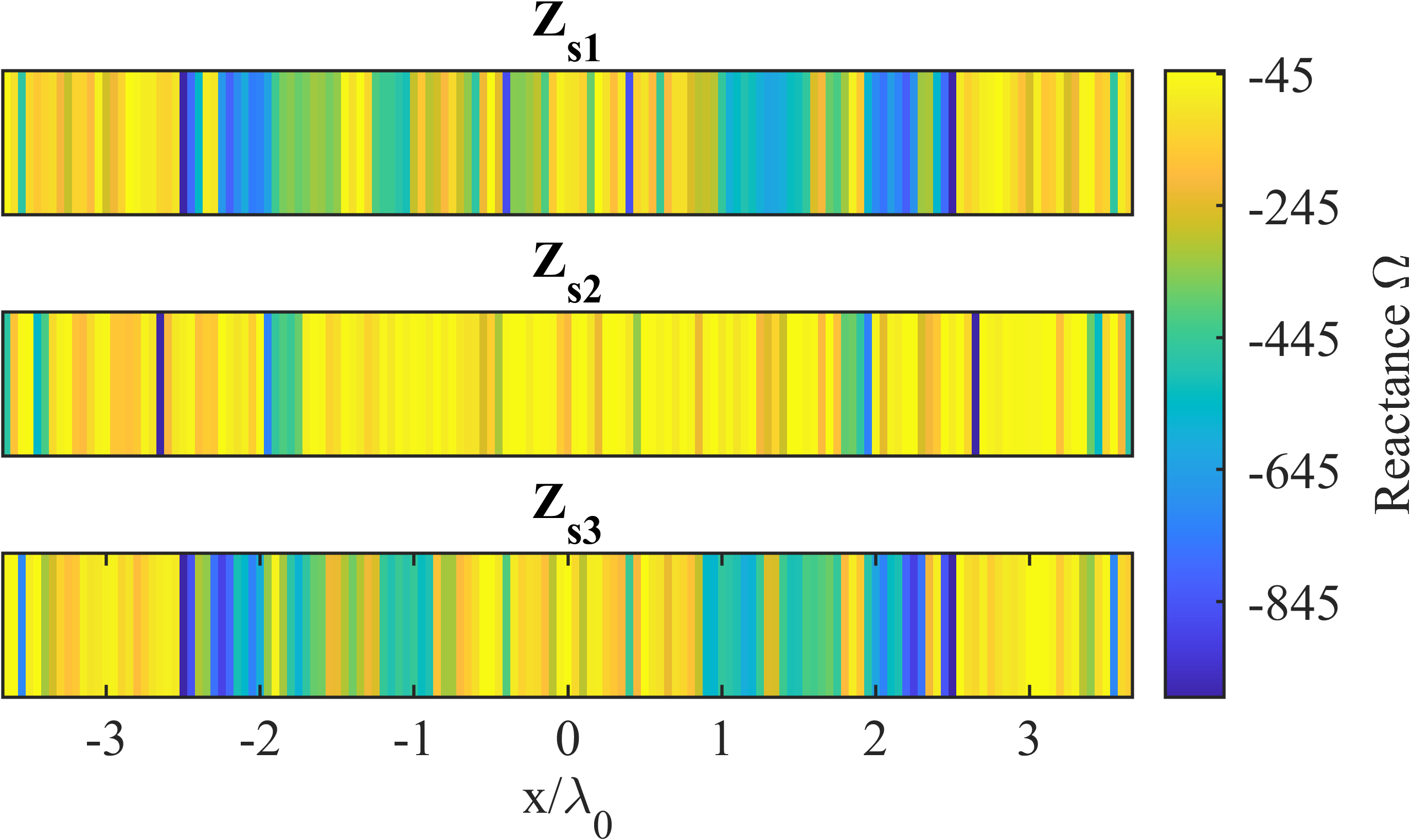}}
\caption{The impedance profiles for the metasurface designed to implement a beam scanned to a 35 degree angle. }
\label{fig9}
\end{figure}

To find the required $E_0$ for full transmission and a lossless metasurface,  (\ref{E_0}) is used. The optimized impedance profiles for the three sheets are shown in Fig. \ref{fig9}.

\begin{figure}[!t]
\subfloat[]{\includegraphics[]{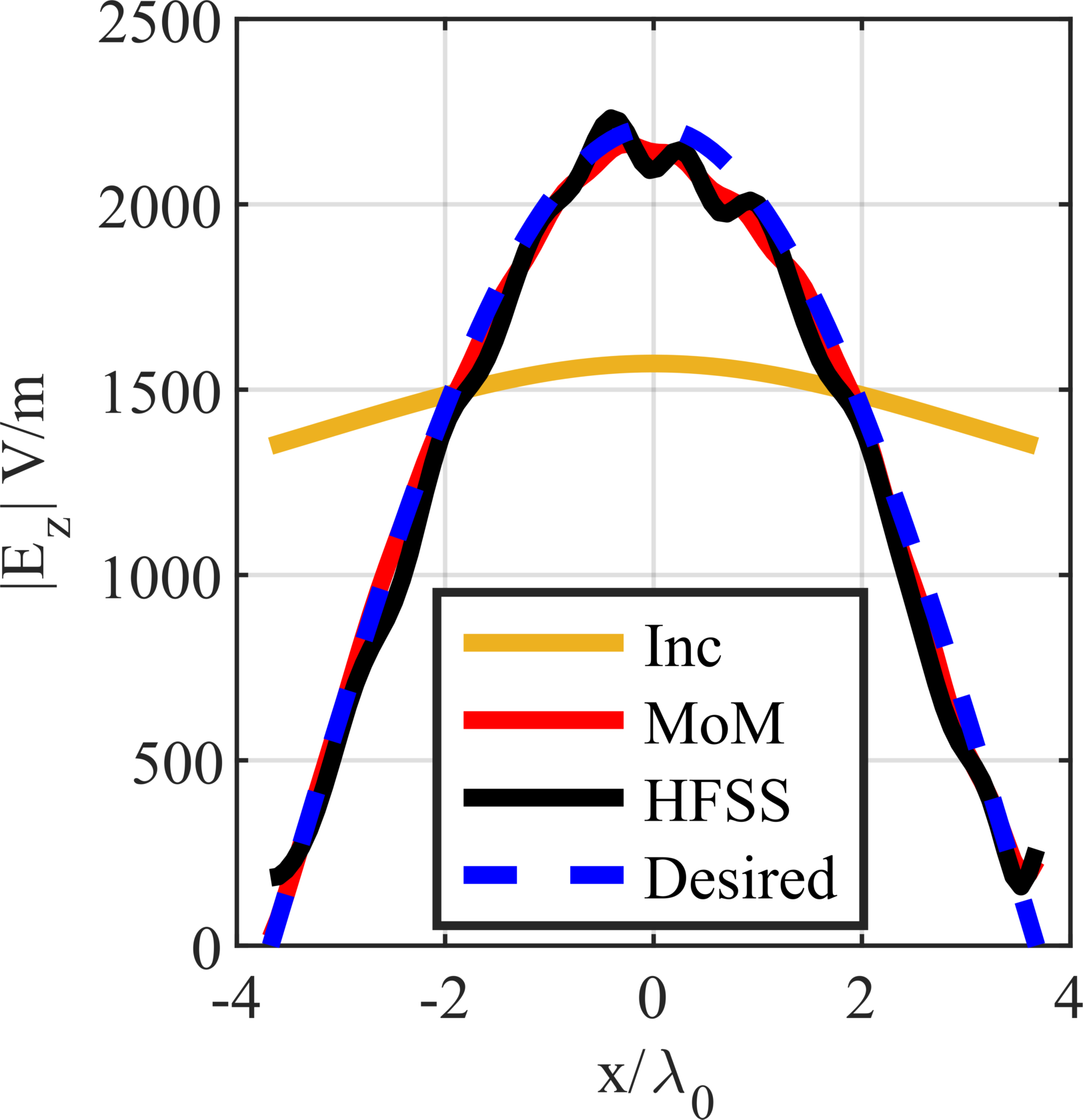}}\hfill
	\subfloat[]{\includegraphics[]{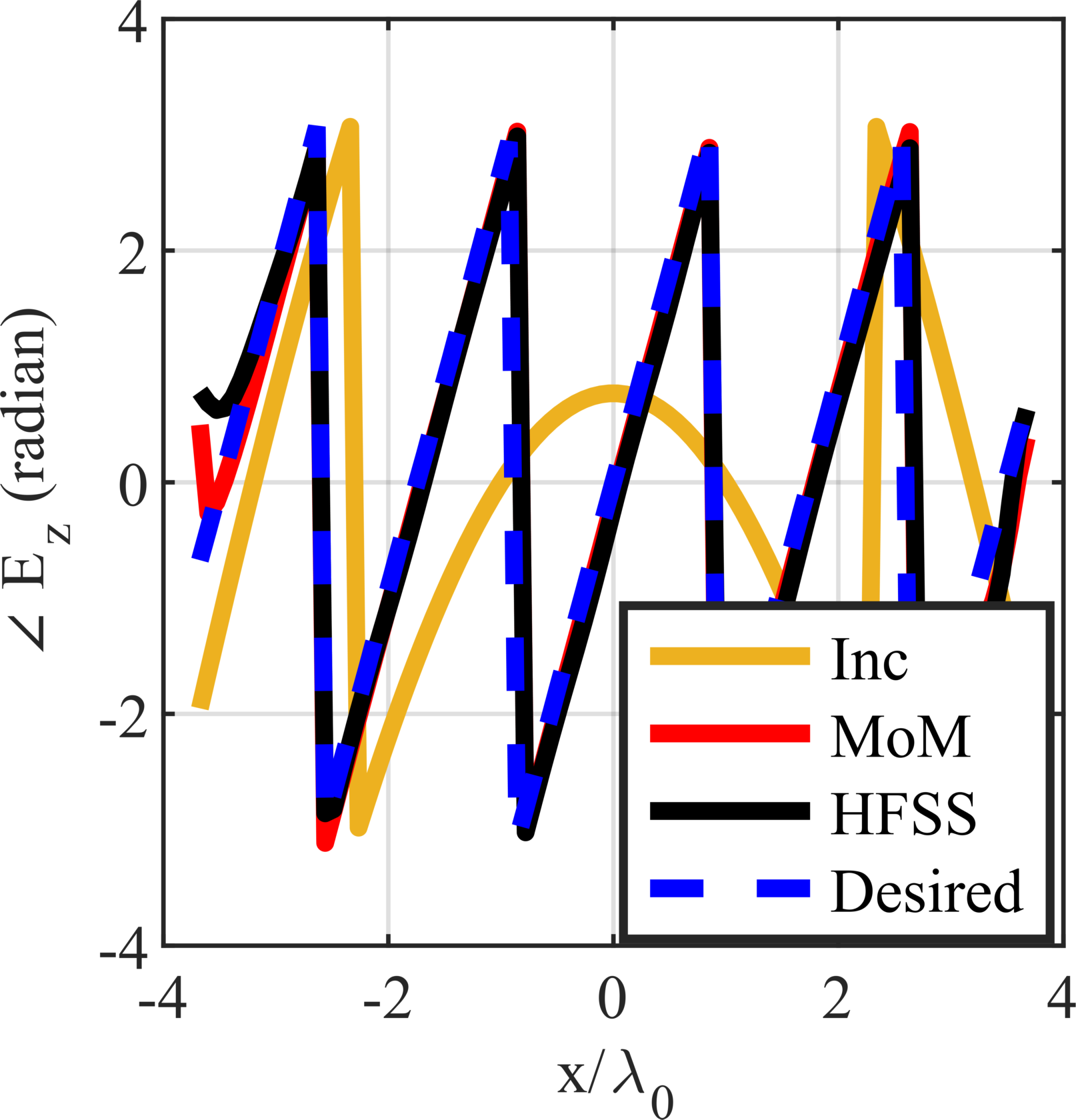}}\\
	\caption{The total electric field transmitted by the metasurface for example III. The incident field is shown in yellow, the desired field in dashed blue, the transmitted field computed using MoM in red, and that computed using HFSS is in black. (a) Electric field amplitude. (b) Electric field phase.   }
	\label{fig10}
\end{figure}

\begin{figure}[!t]
\centerline{\includegraphics[scale=1]{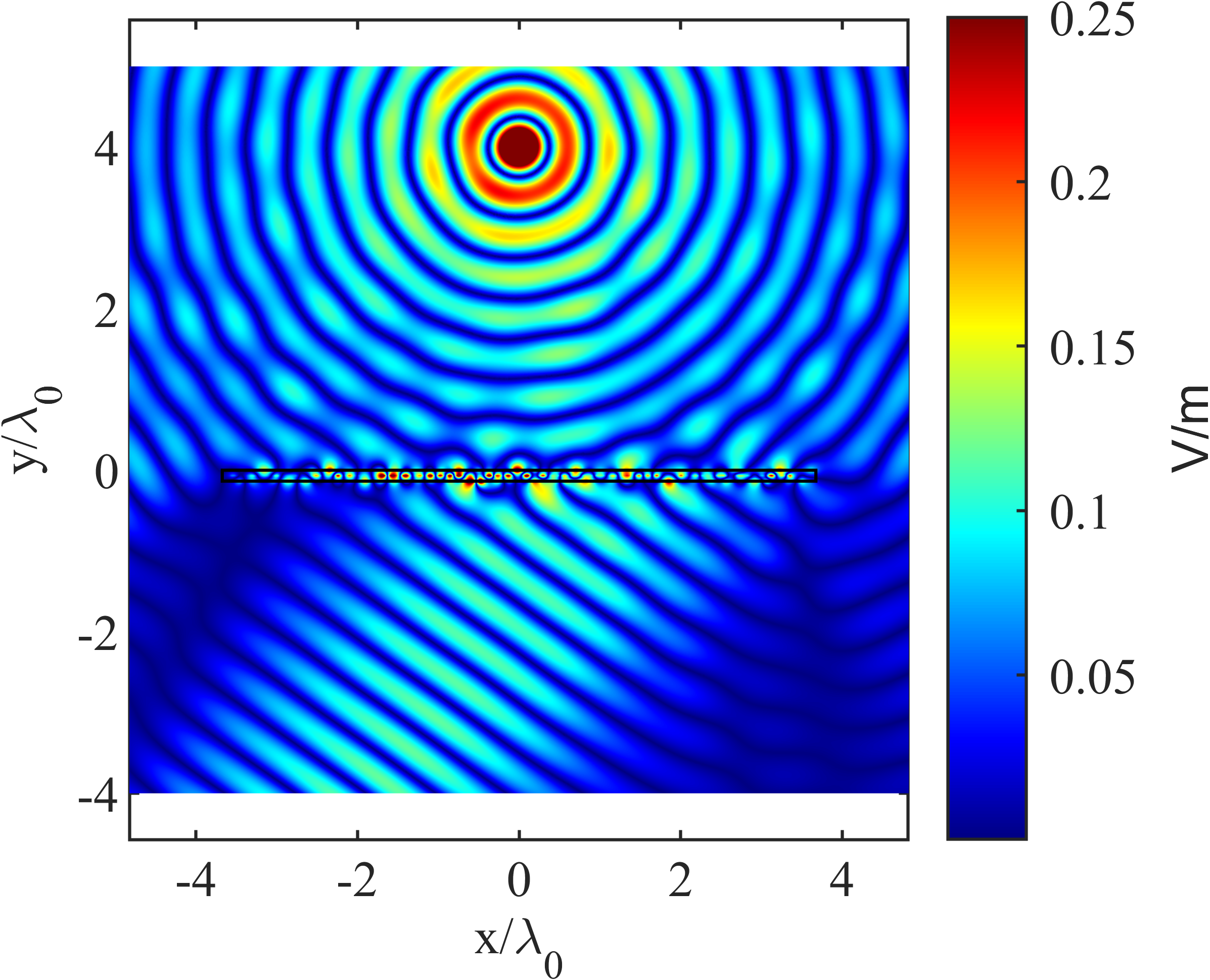}}
\caption{A time snapshot of the electric field from an HFSS full wave simulation for example III. The electric field is normalized to 25\% of the feed maximum value.     }
\label{fig11}
\end{figure}

The electric field amplitude and phase are shown in Fig. \ref{fig10}. The metasurface clearly produces the desired phase distribution and the cosine taper. The electromagnetics full-wave simulation closely agrees with the MoM results. An electric field time snapshot using Ansys HFSS is shown in Fig. \ref{fig11}.

\section{Conclusion}

The accurate design of transmissive metasurfaces using integral equations is reported. The metasurfaces consist of three spatially-varying impedance sheets separated by two dielectric substrates. The metasurface controls the phase and amplitude of the transmitted field, and can be used for beamforming applications. The design method overcomes the issues associated with the commonly used local periodicity approximation. It is explicitly shown that the local periodicity approach produces inaccurate results. A comparison between the two design methods is presented for uniform phase and amplitude. In addition, three design examples are presented that show the ability to control the amplitude and phase of the transmitted field.

%




\end{document}